\documentclass[dvips]{article}
\usepackage{supertabular,lscape,epsfig}
\usepackage{amssymb}
\usepackage{amsmath}
\usepackage{rotating}
\DeclareSymbolFont{ppa}{OT1}{ppl}{m}{it}
\DeclareMathSymbol{\vv}{\mathalpha}{ppa}{'166}

\begin{document}

\newcommand{\dd}{\,{\rm d}}
\newcommand{\ie}{{\it i.e.},\,}
\newcommand{\etal}{{\it et al.\ }}
\newcommand{\eg}{{\it e.g.},\,}
\newcommand{\cf}{{\it cf.\ }}
\newcommand{\vs}{{\it vs.\ }}
\newcommand{\zdot}{\makebox[0pt][l]{.}}
\newcommand{\up}[1]{\ifmmode^{\rm #1}\else$^{\rm #1}$\fi}
\newcommand{\dn}[1]{\ifmmode_{\rm #1}\else$_{\rm #1}$\fi}
\newcommand{\upd}{\up{d}}
\newcommand{\uph}{\up{h}}
\newcommand{\upm}{\up{m}}
\newcommand{\ups}{\up{s}}
\newcommand{\arcd}{\ifmmode^{\circ}\else$^{\circ}$\fi}
\newcommand{\arcm}{\ifmmode{'}\else$'$\fi}
\newcommand{\arcs}{\ifmmode{''}\else$''$\fi}
\newcommand{\MS}{{\rm M}\ifmmode_{\odot}\else$_{\odot}$\fi}
\newcommand{\RS}{{\rm R}\ifmmode_{\odot}\else$_{\odot}$\fi}
\newcommand{\LS}{{\rm L}\ifmmode_{\odot}\else$_{\odot}$\fi}
\newcommand{\feh}{\hbox{$ [{\rm Fe}/{\rm H}]$}}

\newcommand{\Abstract}[2]{{\footnotesize\begin{center}ABSTRACT\end{center}
\vspace{1mm}\par#1\par
\noindent
{~}{\it #2}}}

\newcommand{\TabCap}[2]{\begin{center}\parbox[t]{#1}{\begin{center}
  \small {\spaceskip 2pt plus 1pt minus 1pt T a b l e}
  \refstepcounter{table}\thetable \\[2mm]
  \footnotesize #2 \end{center}}\end{center}}

\newcommand{\TableSep}[2]{\begin{table}[p]\vspace{#1}
\TabCap{#2}\end{table}}

\newcommand{\FigCap}[1]{\footnotesize\par\noindent Fig.\  %
  \refstepcounter{figure}\thefigure. #1\par}

\newcommand{\TableFont}{\footnotesize}
\newcommand{\TableFontIt}{\ttit}
\newcommand{\SetTableFont}[1]{\renewcommand{\TableFont}{#1}}

\newcommand{\MakeTable}[4]{\begin{table}[htb]\TabCap{#2}{#3}
  \begin{center} \TableFont \begin{tabular}{#1} #4
  \end{tabular}\end{center}\end{table}}

\newcommand{\MakeTableSep}[4]{\begin{table}[p]\TabCap{#2}{#3}
  \begin{center} \TableFont \begin{tabular}{#1} #4
  \end{tabular}\end{center}\end{table}}

\newenvironment{references}%
{
\footnotesize \frenchspacing
\renewcommand{\thesection}{}
\renewcommand{\in}{{\rm in }}
\renewcommand{\AA}{Astron.\ Astrophys.}
\newcommand{\AAS}{Astron.~Astrophys.~Suppl.~Ser.}
\newcommand{\ApJ}{Astrophys.\ J.}
\newcommand{\ApJS}{Astrophys.\ J.~Suppl.~Ser.}
\newcommand{\ApJL}{Astrophys.\ J.~Letters}
\newcommand{\AJ}{Astron.\ J.}
\newcommand{\IBVS}{IBVS}
\newcommand{\PASJ}{PASJ}
\newcommand{\PASP}{P.A.S.P.}
\newcommand{\Acta}{Acta Astron.}
\newcommand{\MNRAS}{MNRAS}
\renewcommand{\and}{{\rm and }}
\section{{\rm REFERENCES}}
\sloppy \hyphenpenalty10000
\begin{list}{}{\leftmargin1cm\listparindent-1cm
\itemindent\listparindent\parsep0pt\itemsep0pt}}%
{\end{list}\vspace{2mm}}

\def\TYLDA{~}
\newlength{\DW}
\settowidth{\DW}{0}
\newcommand{\dw}{\hspace{\DW}}

\newcommand{\refitem}[5]{\item[]{#1} #2%
\def\REFARG{#3}\ifx\REFARG\TYLDA\else, {\it#3}\fi
\def\REFARG{#4}\ifx\REFARG\TYLDA\else, {\bf#4}\fi
\def\REFARG{#5}\ifx\REFARG\TYLDA\else, {#5}\fi.}

\newcommand{\Section}[1]{\section{#1}}
\newcommand{\Subsection}[1]{\subsection{#1}}
\newcommand{\Acknow}[1]{\par\vspace{5mm}{\bf Acknowledgments.} #1}
\pagestyle{myheadings}

\newfont{\bb}{ptmbi8t at 12pt}
\newcommand{\xrule}{\rule{0pt}{2.5ex}}
\newcommand{\xxrule}{\rule[-1.8ex]{0pt}{4.5ex}}
\def\thefootnote{\fnsymbol{footnote}}

\begin{center}
{\Large\bf Classical Cepheids in the Milky Way}
\vskip1cm
{\bf
P.~~P~i~e~t~r~u~k~o~w~i~c~z$^1$,~~I.~~S~o~s~z~y~\'n~s~k~i$^1$,~~and~~A.~~U~d~a~l~s~k~i$^1$\\}
\vskip3mm
{
$^1$ Astronomical Observatory, University of Warsaw, Al. Ujazdowskie 4, 00-478 Warszawa, Poland\\
}
\end{center}

\Abstract{We share the most up-to-date, carefully verified list of classical
Cepheids residing in the Galaxy. Based on long-term OGLE experience in
the field of variable stars, we have inspected candidates for Cepheids from
surveys such as ASAS, ASAS-SN, ATLAS, Gaia, NSVS, VVV, WISE, ZTF, among
others, and also known sources from the General Catalogue of Variable Stars.
Only objects confirmed in the optical range as classical Cepheids are included
in the list. We provide Gaia EDR3 identifications of the stars. Purity of the
sample exceeds 97 per cent, while its completeness is of about 88 per cent
down to a magnitude $G=18$. The list contains 3352 classical Cepheids, of which
2140 stars are fundamental-mode pulsators. Basic statistics and comparison
between the classical Cepheids from the Milky Way, Andromeda Galaxy (M31),
and Magellanic Clouds are provided. The list is available at the OGLE Internet
Data Archive.}

{Stars: variables: Cepheids -- Stars: oscillations -- Galaxy: bulge -- Galaxy: disk -- Catalogs}


\Section{Introduction}

Classical Cepheids are extremely important stellar objects for modern
astrophysics (\eg Pilecki \etal 2021) and cosmology (\eg Riess \etal 2019,
Freedman \etal 2021). These pulsating variable stars serve as distance
indicators in the Local Volume (\eg Bono \etal 2010, Scowcroft \etal 2013)
as well as probes for asteroseismology (\eg Pietrzy\'nski \etal 2010,
Smolec and Moskalik 2010, Anderson \etal 2016). Classical Cepheids have
been used to map young components of the Milky Way (\eg Dambis \etal 2015,
Skowron \etal 2019a) and nearby galaxies such as the Large Magellanic
Cloud (\eg Jacyszyn-Dobrzeniecka \etal 2016, Inno \etal 2016). Follow-up
observations of the Milky Way's Cepheids have allowed for the determination
of metallicity gradients (Genovali \etal 2015) and for precise
drawing the rotation curve of the outer Galactic disk (Mr\'oz \etal 2019).

Classical Cepheids are Population I stars of luminosity classes from
Ia (luminous supergiants) through Iab (normal supergiants) and Ib
(underluminous supergiants) to II (bright giants) and spectral types from
F5 to K2. In the Hertzsprung-–Russell diagram, the stars lie on the
classical instability strip. Classical Cepheids pulsating in the fundamental
mode (F) show light variations with periods from about 1~day to over 200~days.
Their light curve shapes are usually asymmetric with a rapid rise and a slower
decline. In the period range from 5 to 23 days, they are characterized by the
presence of an additional bump at location depending on the pulsation period.
This effect is known as the Hertzsprung progression and seems to be caused by
the 2:1 resonance between the fundamental mode and the second overtone (2O).
Light curves of fundamental-mode Cepheids show sharp extrema and have
$I$-band amplitudes up to about 1.0~mag (or 1.7~mag in $V$). First-overtone
(1O) classical Cepheids have pulsation periods in the range from about
0.23 day up to almost 10 days. They have smoother light curves, round
minima, and on average twice lower amplitudes than those in fundamental-mode
stars. Some of the observed classical Cepheids pulsate in two or three radial
modes simultaneously. Multi-mode Cepheids have characteristic period ratios
which helps to distinguish them from multi-mode RR Lyr-type stars
(\eg Smolec \etal 2017). In some first-overtone Cepheids, additional
periodicities corresponding to high-order non-radial modes have been
discovered (Soszy\'nski \etal 2008, Moskalik and Ko{\l}aczkowski 2009,
Rathour \etal 2021). Light curves of Cepheids are very stable over time.
Only several Galactic classical Cepheids show definite amplitude and phase
modulations, in particular Polaris ($\alpha$~UMi; Bruntt \etal 2008)
and V473 Lyr (Moln\'ar and Szabados 2014).

The main purpose of the work here is presentation of the most up-to-date
and as pure as possible list of Galactic classical Cepheids
that may serve in various astrophysical studies. We plan to update
the list once new classical Cepheids are discovered. In this paper,
we analyse the whole set of known Milky Way's Cepheids and compare it
with collections of such stars residing in M31 and Magellanic Clouds.


\Section{Discovering Classical Cepheids}

The first classical Cepheids, $\eta$~Aql and $\delta$~Cep (the prototype
object of the whole class), were discovered by Edward Pigott and John Goodricke
in 1784, respectively. Thirty new Cepheids were found over a century following
that discovery. In the first edition of the General Catalogue of Variable Stars
(GCVS), there were 473 objects identified as ``long-period Cepheids''
(Kukarkin and Parenago 1948)\footnote{A term ``short-period Cepheids''
referred to saw-tooth-shaped variables with periods shorter than a day observed
mainly in globular clusters and nowadays classified as RR Lyr-type stars.}.
Twenty-eight of the stars were marked as uncertain cases. The most recent
version (from June 2021) of the fifth GCVS edition (Samus \etal 2017)
contains 632 classical Cepheids identified as fundamental-mode (DCEP),
first-overtone (DCEPS), and double-mode (beat) pulsators (DCEP(B)).
In the catalog, there are also 224 variable sources classified to one
of the following ambiguous or uncertain types (colon extension):
CEP, CEP(B), CEP:, DCEP:, and DCEPS:. Some of the sources are type II
(or Population II) Cepheids. Some classical Cepheids listed in GCVS
belong to the Magellanic Clouds.

The number of known Galactic classical Cepheids has grown considerably in
the era of massive wide-field photometric CCD surveys. One of the first such surveys
was the All Sky Automated Survey (ASAS; Pojmanski 1997) operated from the Las
Campanas Observatory, Chile, since 1996. ASAS-3 has monitored stars down to
$V\approx14.5$~mag over the entire southern sky up to declination
$\delta\approx+28\arcd$. The on-line ASAS Catalog of Variable Stars
(ACVS; Pojmanski 2002, 2003, Pojmanski and Maciejewski 2004, 2005,
Pojmanski \etal 2005) contains 870 objects classified as authentic classical
Cepheids (labeled as DCEP-FU, DCEP-FO, DCEP-FU/DCEP-FO or DCEP-FO/DCEP-FU)
and 798 objects as possible classical Cepheids (with uncertain classification
or multiple identifications).

A very similar robotic project was carried out from Los Alamos, New Mexico,
USA, by the Northern Sky Variability Survey (NSVS; Wo\'zniak \etal 2004).
The survey monitored the entire northern hemisphere and also southern fields
down to $\delta\approx-38\arcd$ (although with fewer epochs) over one
full year (from April 1999 to March 2000). In contrast to ASAS, which
used the Johnson $V$ and Cousins $I$ filters, NSVS collected
unfiltered photometric data. Hoffman \etal (2009) performed an automated
classification of NSVS stars and identified several hundreds of
long-period variables including Cepheids. Due to photometric constraints,
they were not able to distinguish between classical and type II Cepheids,
but nearly one hundred stars were cross-matched with already known classical
Cepheids from GCVS. Over a decade several additional objects from the NSVS
database were identified as classical Cepheids by various researchers (\eg
Benk\"o and Csubry 2007, Kuzmin 2008, Khruslov and Kusakin 2016).

Due to a large pixel scale ($\approx15\arcs$), the ASAS and NSVS
instruments were insufficient to resolve stars in dense regions close
to the Galactic plane, where the Population I Cepheids reside.
These regions have been extensively observed by the Optical
Gravitational Lensing Experiment (OGLE) using the 1.3-m Warsaw
telescope located at the Las Campanas Observatory since 1997.
The third phase of the project (OGLE-III, Udalski \etal 2008), conducted
in the years 2001--2009, provided the identification of 32 classical Cepheids
in the Galactic bulge (Soszy\'nski \etal 2011) and 20 classical Cepheids
in a small part of the Galactic disk (Pietrukowicz \etal 2013). A new wide-field
camera installed in 2010 started the fourth phase of the project (OGLE-IV,
Udalski \etal 2015) and allowed for a regular deep monitoring of the entire
Galactic stripe seen from Las Campanas. A total of 1935 classical Cepheids
were reported based on OGLE-IV time-series data (Soszy\'nski \etal 2017,
Udalski \etal 2018, Skowron \etal 2019b, Soszy\'nski \etal 2020).
Four additional classical Cepheids were found in the OGLE-II data
(from years 1997--2000) by three amateur astronomers: R.~Jansen in 2009,
S.~H\"ummerich in 2013, and J.~Falc\'on Quintana in 2017.
The OGLE observations are collected mainly in the $I$ band and reach
$I\approx21.5$~mag for the inner Galactic bulge and $I\approx20.5$~mag
for the Galactic disk and outer bulge. Regular, long-term, high-quality
OGLE photometry, including data for Magellanic Cloud Cepheids
(\eg Soszy\'nski \etal 2008, 2015), has allowed us to verify and
properly classify new candidates for Galactic Cepheids from OGLE
as well as from other surveys. The OGLE collection of periodic variable stars
currently contains over a million objects.

In the 2010s several new large-scale surveys, focused not only on variable star
research, started to operate. After 2--3 years of observations the survey
teams released their first time-series data and shortly thousands of variable
sources and hundreds of candidates for Cepheids were published.
One of such surveys, called Asteroid Terrestrial-impact Last Alert System
or ATLAS (Tonry \etal 2018), employs two 0.5-m Schmidt telescopes at the Hawaii
Haleakala Observatory since 2015 and Mauna Loa Observatory since 2017.
For the asteroid search the system use non-standard broad band filters,
cyan ($c$; 420–-650 nm) and orange ($o$; 560–-820 nm). Heinze \etal (2018)
reported on the detection of 25~162 candidate pulsating variables with
saw-tooth-shaped light curves, including 1140 candidates for Cepheids
down to $c\approx19$ mag, $o\approx18.5$ mag, or $r\approx18.5$ mag found
between declination $-30\arcd$ and $+60\arcd$.

Another large-scale survey, the All-Sky Automated Survey for Supernovae (ASAS-SN)
started with the installation of the first station at the Haleakala Observatory
in Hawaii in 2013. One year later the project was extended for the southern
station at the Cerro Tololo Inter-American Observatory, Chile. ASAS-SN
patrols the entire sky for supernovae and various transients to a depth
of $V\approx17$ mag (Shappee \etal 2014, Kochanek \etal 2017). In June 2021
the survey database contained information on 666~502 variable stars, of which
2185 objects were classified as classical Cepheids (Jayasinghe \etal 2018, 2019, 2020).
Some of the Cepheids are members of the Large Magellanic Cloud (LMC)
and Small Magellanic Cloud (SMC). In the presented here work, the
long-term ASAS and ASAS-SN data were particularly useful in the verification
of known and recently detected candidates for pulsating stars in less dense
regions at higher Galactic latitude.

A deep optical monitoring (down to $g,r\approx21$~mag, $i\approx20.5$~mag) of the
northern part of the sky ($\delta > -25\arcd$) is conducted by the Zwicky Transient
Facility (ZTF; Bellm \etal 2019) using the 1.2-m Samuel Oschin telescope at the
Palomar Observatory, USA, since 2018. The second data release of the survey (ZTF DR2)
allowed for the detection of 781~602 periodic variable stars including 582
candidates for classical Cepheids (Chen \etal 2020). Over 20 long-period Cepheids
are stars from M31.

Highly extincted regions close to the Galactic plane are practically inaccessible
for optical surveys. In recent years two surveys have obtained near-infrared
time-series photometry in the area of the Galactic bulge and adjacent sections
of the Galactic disk. Since 2001 the SIRIUS camera installed on the InfraRed
Survey Facility (IRSF/SIRIUS) 1.4-m telescope at the South African Astronomical
Observatory has collected series of $JHK_s$ images of selected fields near the
Galactic equator (Matsunaga \etal 2011). A result of the searches was the
detection of about 30 candidates for classical Cepheids among hundreds of
long-period ($<60$~d) variable stars (Matsunaga \etal 2016, Tanioka \etal 2017).
The other near-infrared survey, VISTA Variables in the Via Lactea (VVV;
Minniti \etal 2010), is an ESO public survey conducted on the 4.1-m VISTA
telescope located at the Paranal Observatory, Chile. The multi-epoch $K_s$-band
and single-epoch $ZYJH$-band observations obtained in years 2010--2019 have
brought the detection of 689 candidates for classical Cepheids in the
galactic longitude range $-65\zdot\arcd3 < l < +10\zdot\arcd4$ and latitude range
$-2\zdot\arcd25\arcd < b < +2\zdot\arcd25\arcd$ (D\'ek\'any \etal 2015ab, 2019).

Since 2014 the whole sky is repeatedly scanned from space by Gaia astrometric
mission. Gaia photometry is collected in white-light $G$-band (330-–1050 nm)
and reaches $G\approx21$~mag. Based on the second data release (Gaia DR2,
Gaia Collaboration \etal 2018) Clementini \etal (2019) informed about 350
candidates for new classical Cepheids. Other space mission which data have been used
to search for periodic variables is Wide-field Infrared Survey Explorer (WISE).
Chen \etal (2018) reported on the detection of 1312 candidates for Cepheids.
However, a weak side of WISE is high angular resolution of its camera
and clumping of the time-series observations.

Data from several other surveys have been searched for pulsating
stars which led to the discovery of some 20 new classical Cepheids.
The surveys were: the Galactic Spiral Arms program in the second phase
of Exp\'erience de Recherche d'Objets Sombres (EROS2-GSA; Derue \etal 2002),
the Berlin Exoplanet Search Telescope (BEST; Kabath \etal 2008, 2009),
the Multitudinous Image-based Sky-survey and Accumulative Observations
(MISAO)\footnote{http://www.aerith.net/misao/index.html},
Dauban Survey\footnote{http://www.aspylib.com/newsurvey/},
Two Micron All Sky Survey (2MASS; Quillen \etal 2014), Catalina Sky Survey
(CSS; Drake \etal 2014), the Super Wide Angle Search for Planets (SuperWASP)
program\footnote{http://www.superwasp.org/}, and the Convection, Rotation
and planetary Transits (CoRoT) space mission (Poretti \etal 2015). Candidates for
classical Cepheids were also found during dedicated searches for variable objects
in open clusters (\eg Clark \etal 2015). Turner \etal (2009) informed about a
small-amplitude Cepheid that displays remarkably similar properties to Polaris.
Discoveries of several classical Cepheids based on NSVS or SuperWASP photometry
were reported directly to the Variable Star Index (VSX, Watson \etal 2006)
by A.~Prokopovich in 2011, A.~Ditkovsky in 2015, I.~Sergey in 2015,
and M.~Bajer in 2018.


\Section{The List of Galactic Classical Cepheids}

With this paper, we share the most up-to-date and carefully verified
list of Galactic classical Cepheids. This is a continuation of the
work presented in Pietrukowicz \etal (2013) when the list included
841 objects. Since then, several surveys have searched the sky for
Cepheids and published their results. In the verification process,
we visually inspected the light curves and checked various observational
properties of the stars. Nearly all faint objects ($10<I<20$ mag)
come from two surveys, OGLE (southern sky) and ZTF (northern sky).
Brighter objects ($7<V<17$ mag) were inspected thanks to long-term $V$-band
photometry from the ASAS and ASAS-SN surveys. In the case of about 40
brightest classical Cepheids in the sky, we downloaded $BVI$ light curves
obtained by the American Association of Variable Star Observers
(AAVSO)\footnote{https://www.aavso.org/data-download}. We cleaned
the light curves from evident outlying points and corrected the periods
using the T{\footnotesize ATRY} code (Schwarzenberg-Czerny 1996).

The most important criteria used in the verification process were the
characteristic shape and stability of the light curve. In ambiguous
cases, we rely on the position of the variable star in the Fourier space
(coefficient combinations $R_{21}$, $R_{31}$, $\phi_{21}$, and $\phi_{31}$
\vs pulsation period), its amplitude value, position in the sky,
position in the color-magnitude diagram, and spectral type, if were available.
In contrast to the fundamental-mode pulsators, first-overtone pulsators are
more difficult to confirm, in particular in the period range from 2--3 days.
In this range, the light curves have a nearly symmetric, quasi-sinusoidal
shape at low amplitude. In the case of multi-mode pulsators, we checked
their position in the Petersen diagram. A small fraction of variables
detected by ground-based optical surveys like ASAS-SN and ZTF and
classified as Cepheids by automatic classifiers are in fact of
other types, such as eclipsing binaries and rotating (spotted) variables.
Large-amplitude spotted (chromospherically active) variables can
often mimic pulsations, but their light curves clearly change from season
to season (Pietrukowicz \etal 2015). The automatic classifiers rarely
make mistakes in proper classification between classical and type II Cepheids,
but very often put anomalous Cepheids to the group of classical Cepheids.
Anomalous Cepheids as intermediate-age or old objects can be present at high
Galactic latitudes. Long-term ASAS-SN observations show that
some GCVS objects require re-classification.

The list of Galactic classical Cepheids is available to the astronomical
community through the OGLE Internet Data Archive:
\begin{center}
{\it https://www.astrouw.edu.pl/ogle/ogle4/OCVS/allGalCep.listID\\}
\end{center}
In the file, for each Cepheid we provide name of the star together with
the source of the name or survey, equatorial and galactic coordinates based
on Gaia Celestial Reference Frame (Gaia Collaboration \etal 2021), pulsation
mode or modes, period or periods in case of multi-mode pulsators,
identifiers in the OGLE, ASAS, ASAS-SN, Gaia EDR3 catalogs, and mean
$G$-band magnitude. If the Cepheid has a given GCVS designation, this name
is treated as the official one, according to the recommendation of the
International Astronomical Union. If not, we provide the name given
by the discovery survey team. We cross-matched our list of classical
Cepheids with the Gaia EDR3 catalog by adopting a matching radius of
$0\zdot\arcs5$. Over 50 objects without a counterpart were inspected
individually. There is no Gaia EDR3 identifier for four classical Cepheids,
extremely bright Polaris and three faint classical Cepheids toward
the Galactic bulge. There is no $G$-band information for fifteen Cepheids.
According to Gaia Collaboration \etal (2021), the source list
published with Gaia EDR3 in December 2020 will be the same for
Gaia DR3 planned for the first half of 2022.

In Table~1, we summarize the number of Milky Way's classical Cepheids
discovered by various surveys.

\begin{table}[htb!]
\centering \caption{\small Census of known Galactic classical Cepheids with
information on the source of data}
\medskip
{\small
\begin{tabular}{lr}
\hline
Source/Survey & Number \\
\hline
GCVS          &   737 \\
ASAS          &   108 \\
NSVS          &    40 \\
OGLE          &  1690 \\
ATLAS         &   139 \\
ASAS-SN       &   161 \\
WISE          &    12 \\
Gaia-DR2      &     5 \\
VVV           &    21 \\
ZTF-DR2       &   422 \\
other         &    17 \\
\hline
Total         &  3352 \\
\hline
\end{tabular}}\\
\end{table}


\Section{Observational Properties of Classical Cepheids}

In this section, we analyse selected observational properties and
provide some statistics on the Galactic classical Cepheids. In Fig.~1,
we show a map of all known 3352 classical Cepheids in galactic coordinates.
These stars as representatives of the young Milky Way's component (the
thin disk) concentrate around the Galactic plane. In the distribution presented
in Fig.~2, we count the stars in $8\arcd$-wide bins along the galactic
longitude. The lowest number of stars is observed, not surprisingly,
in the direction of the anticenter. Two maxima, the highest one around
$l=-72\arcd$ and a smaller one at $l\approx+48\arcd$, correspond to the
inner edge of the tangent lines to the Carina arm and Sagittarius arm,
respectively (Vall\'ee 2017). Hundreds of unknown classical Cepheids likely
reside between the two tangent lines, but are hidden behind clouds of
dust. There is a local minimum in the Cepheid distribution around
$l=+80\arcd$, also visible in the map. The minimum coincides with
a structure called the Northern Coalsack in constellation Cygnus.
It is a unique area in the Galactic disk worth variability exploration
in near-infrared bands.

\begin{figure}[]
\centerline{\includegraphics[angle=0,width=112mm]{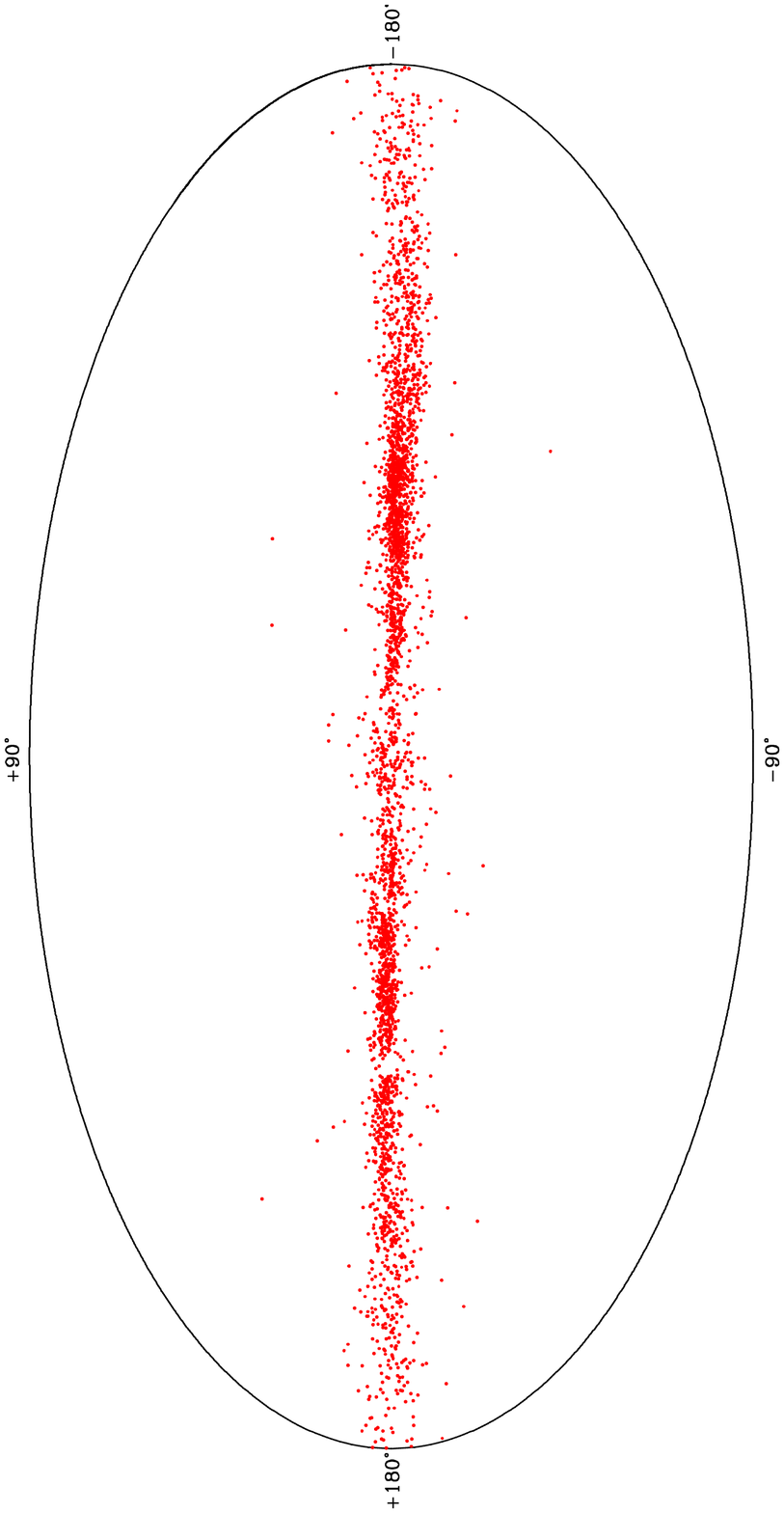}}
\FigCap{Distribution, in Galactic coordinates, of 3352 optically confirmed
classical Cepheids residing in the Milky Way.}
\end{figure}

\begin{figure}[htb!]
\centerline{\includegraphics[angle=0,width=120mm]{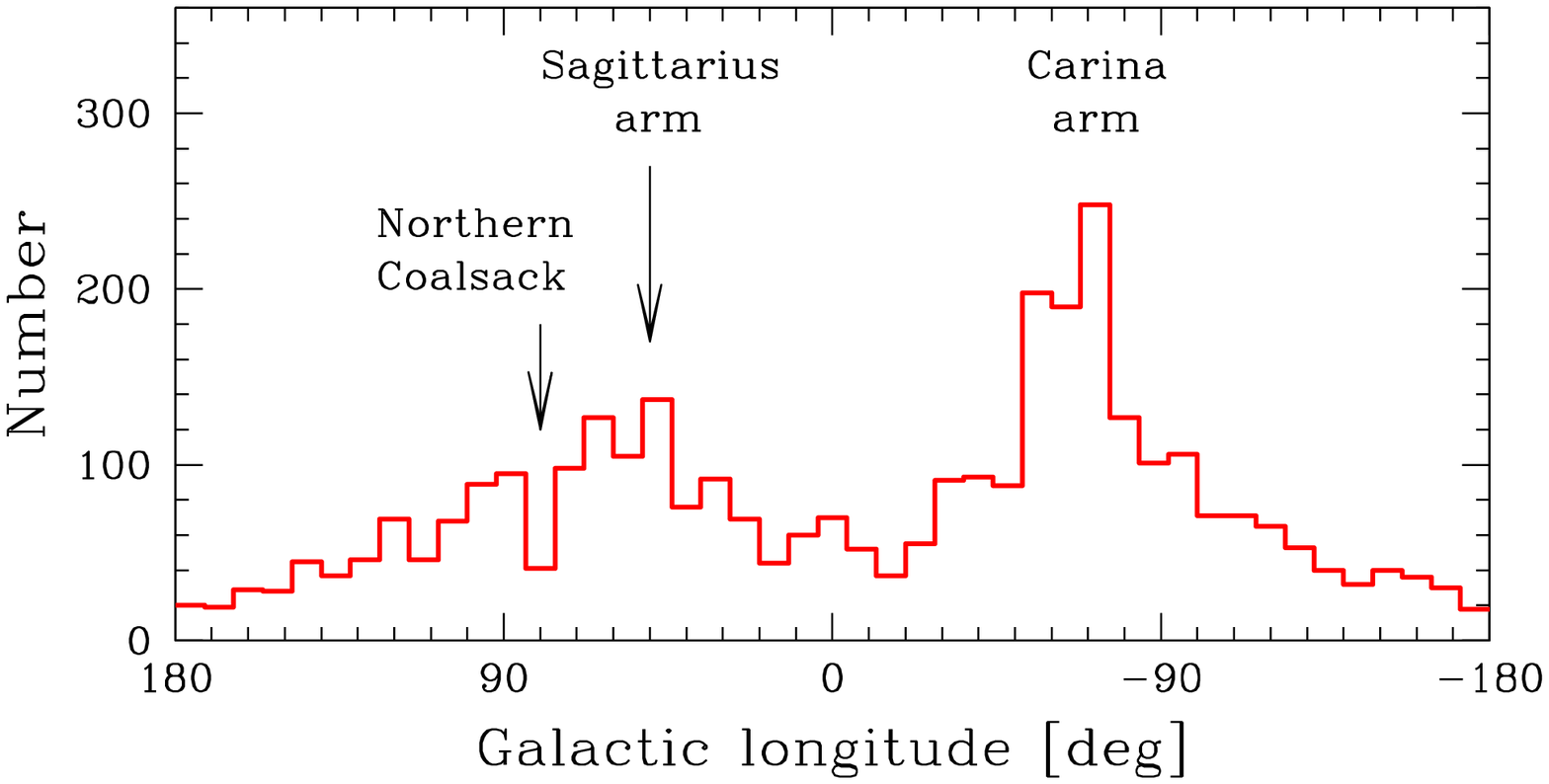}}
\FigCap{Distribution of confirmed classical Cepheids in the function
of galactic longitude. Some extrema in this distribution are
related to known structures.}
\end{figure}

In Fig.~3, we present mean $G$-band brightness distributions for all (but the
mentioned fifteen) known Galactic classical Cepheids and, separately, for
stars pulsating in the fundamental mode only. Both distributions have
the maximum around $G=16$ mag. We divided each of the two sets of stars
into two complementary parts, stars from highly reddened Galactic area
($-80\arcd<l<+50\arcd$, $-2\arcd<b<+2\arcd$) and stars outside this area.
The maximum of the distribution for fundamental-mode Cepheids in the
highly-reddened regions reaches $G\approx18$ mag (red histogram in the
lower panel in Fig.~3). The distribution for fundamental-mode
pulsators from mildly reddened regions (blue histogram in the same panel)
is nearly symmetric around $G=14.5$ mag, which suggests that searches
for Cepheids in those regions are highly complete.

\begin{figure}[htb!]
\centerline{\includegraphics[angle=0,width=120mm]{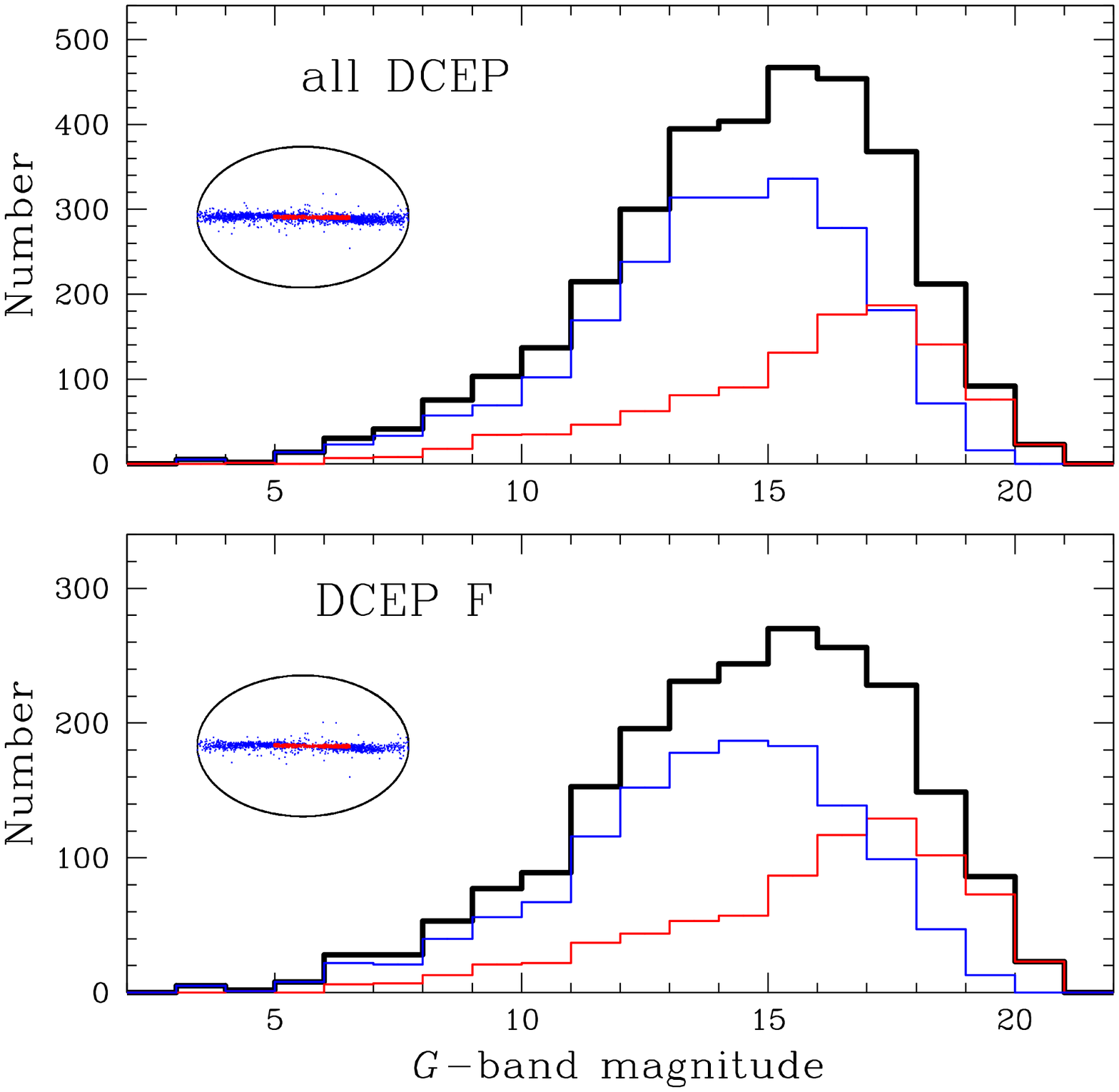}}
\FigCap{Mean $G$-band brightness distributions (black histograms) for all
known ({\it upper panel}) and fundamental-mode Galactic classical Cepheids
({\it lower panel}). In each panel, stars are divided into two complementary groups
depending on their location in the sky: highly obscured regions (red)
and mildly obscured regions (blue), as shown in the inset maps.}
\end{figure}

In Fig.~4, we compare period distributions of fundamental-mode
and first-overtone classical Cepheids from four galaxies in the Local
Group: the Large and Small Magellanic Clouds (Soszy\'nski \etal 2015),
Milky Way (this work), and M31 (Kodric \etal 2018). The distributions
of fundamental-mode pulsators from the Milky Way (2140 stars) and M31
(1662 stars) look remarkably similar to each other. The differences are
that the maximum of the M31 distribution is slightly shifted toward longer
periods and that there is a notable deficit of Milky Way's Cepheids around
$P\approx10$~d. The number of known M31 first-overtone Cepheids is three
times smaller than the number of Galactic pulsators of this type (307 \vs 924).
Due to low luminosity of short-period Cepheids, the current sample of
M31 overtone Cepheids is far from being complete and its period distribution
starts at a relatively high value, from about $P\approx1.5$~d.
It seems that there are missing first-overtone Cepheids around the maximum
in the distribution for Milky Way or stars with periods of about 3 days. In
contrast to the Milky Way and M31, the collections of classical Cepheids for
nearby Magellanic Clouds are complete. The shapes of the distributions for the
two irregular galaxies are quite similar to each other, but differ from
the distributions for the large spirals. The former are skewed toward
shorter periods. The period shift in the distributions is evident in
the fundamental-mode as well as first-overtone Cepheids. This effect
results from different metallicity of the investigated galaxies---the
higher metal content, the longer period of the maximum distribution.

\begin{figure}[htb!]
\centerline{\includegraphics[angle=0,width=120mm]{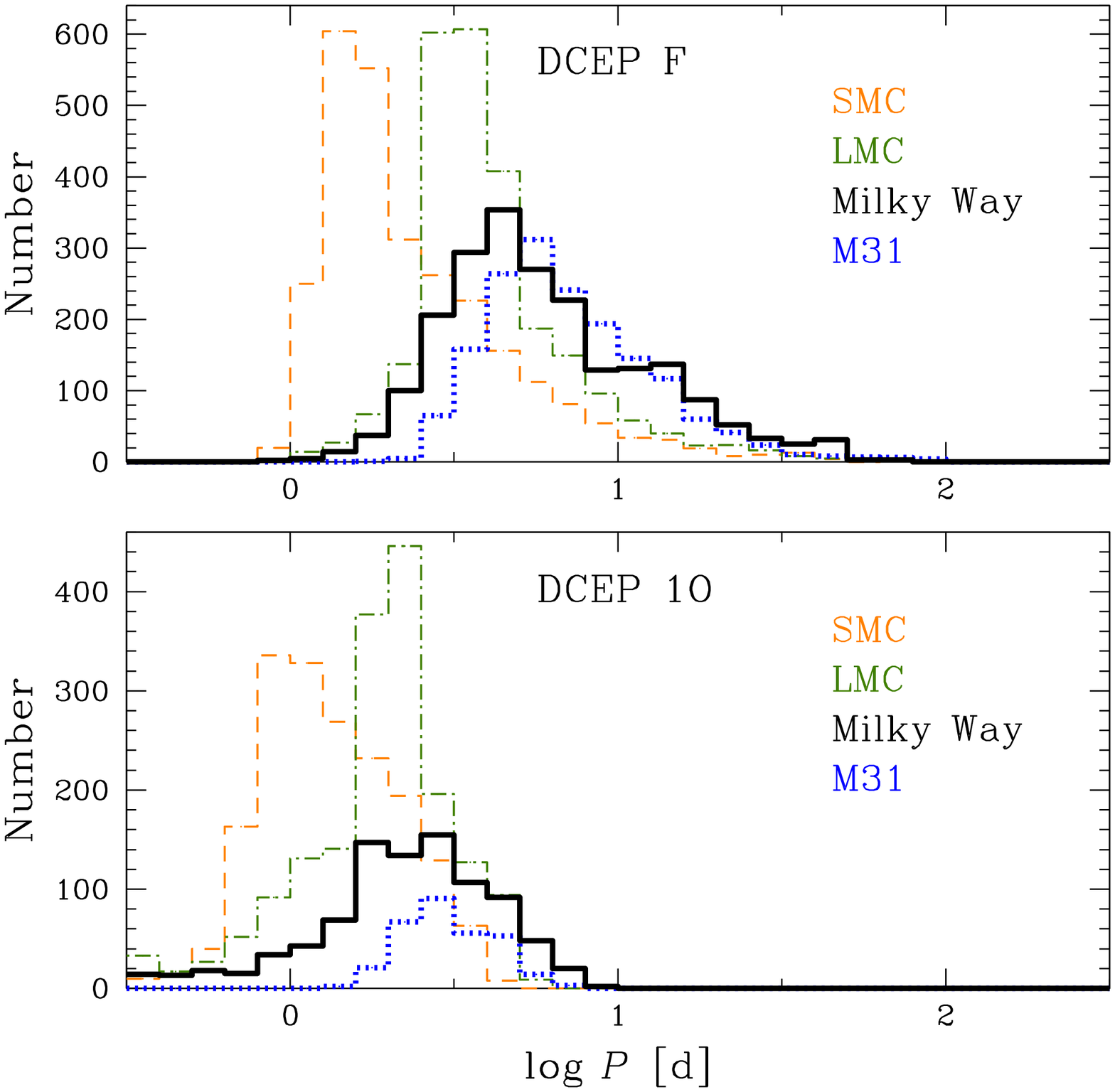}}
\FigCap{Pulsation period distributions for fundamental-mode ({\it upper panel})
and first-overtone classical Cepheids ({\it lower panel}) from the SMC, LMC,
Milky Way, and M31.}
\end{figure}

In Fig.~5, we present a period--amplitude diagram based on $I$-band
observations from the OGLE survey for 1027 fundamental-mode and 540
first-overtone Milky Way's classical Cepheids. Nearly all variables have
full amplitudes higher than 0.1 mag in $I$. There are fundamental-mode
stars with amplitudes reaching 1.0 mag, but most of this type of
pulsators have amplitudes between 0.3 and 0.8 mag. The first-overtone
Cepheids usually pulsate with amplitudes from 0.1 to 0.35 mag, but
may reach 0.6 mag. We cannot exclude that there are more
Cepheids with amplitudes below 0.1 mag.

\begin{figure}[htb!]
\centerline{\includegraphics[angle=0,width=120mm]{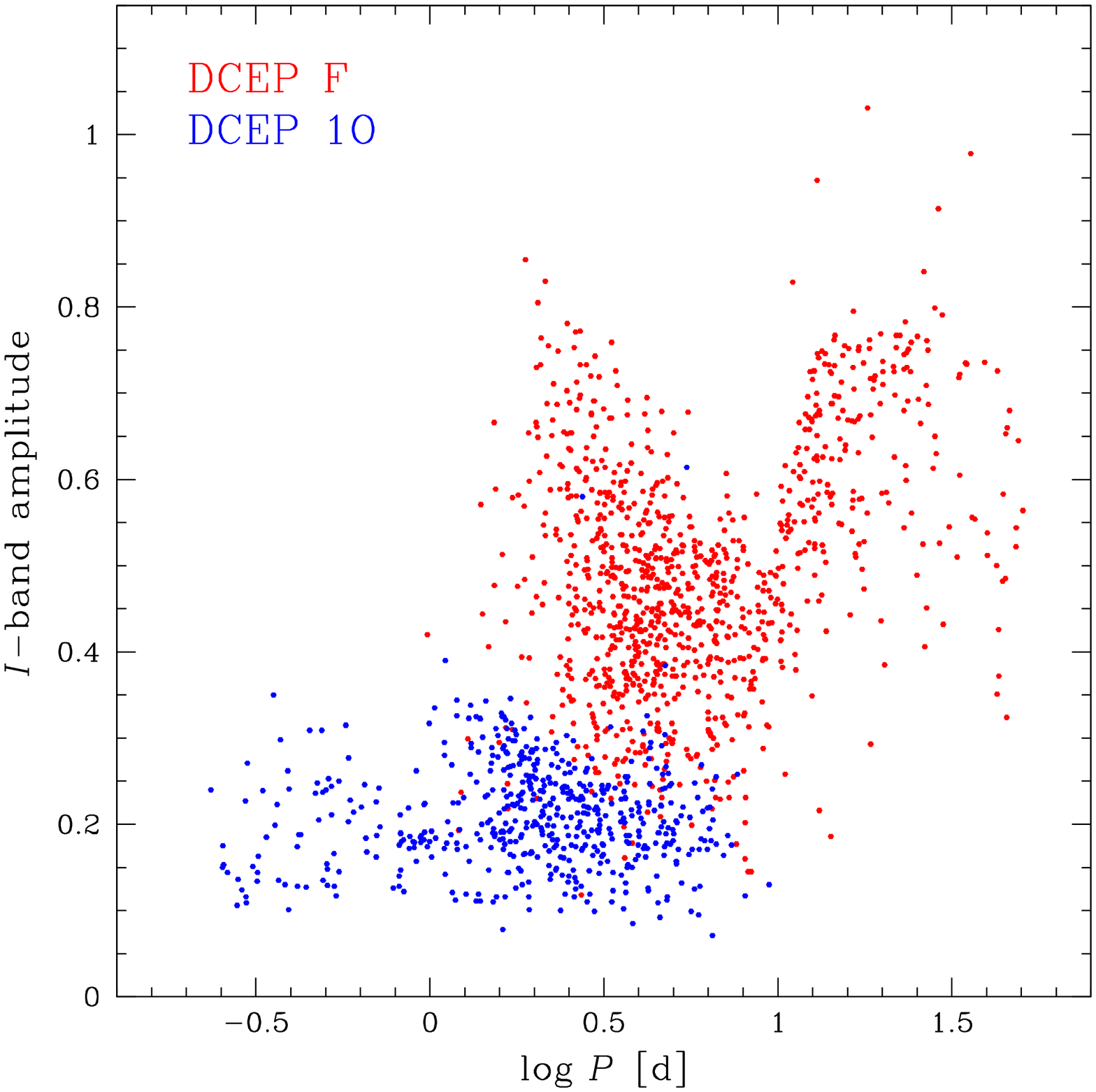}}
\FigCap{Period--amplitude diagram for fundamental-mode (red points)
and first-overtone Galactic classical Cepheids (blue points).}
\end{figure}

Fig.~6 shows where we can find fundamental-mode and first-overtone
classical Cepheids in the Fourier parameter space. The following
Fourier parameter combinations were calculated for $I$-band
light curves from OGLE: $R_{21}=A_2/A_1$, $R_{31}=A_3/A_1$,
$\phi_{21}=\phi_2-2\phi_1$, and $\phi_{31}=\phi_3-3\phi_1$, where $A_i$
and $\phi_i$ are parameters of the cosine Fourier series fitted to the
light curves. Parameters derived from $V$-band observations may differ
slightly. Fundamental-mode Cepheids with periods around 10 days have nearly
symmetric light curves. Similar situation is for first-overtone
Cepheids with periods at approximately 3 days. Around these periods,
the $R_{21}$ and $R_{31}$ ratios are close to zero, $\phi_{21}$
and $\phi_{31}$ combinations dramatically change around zero,
and amplitudes of the stars are lower (see Fig.~5).

\begin{figure}[htb!]
\centerline{\includegraphics[angle=0,width=120mm]{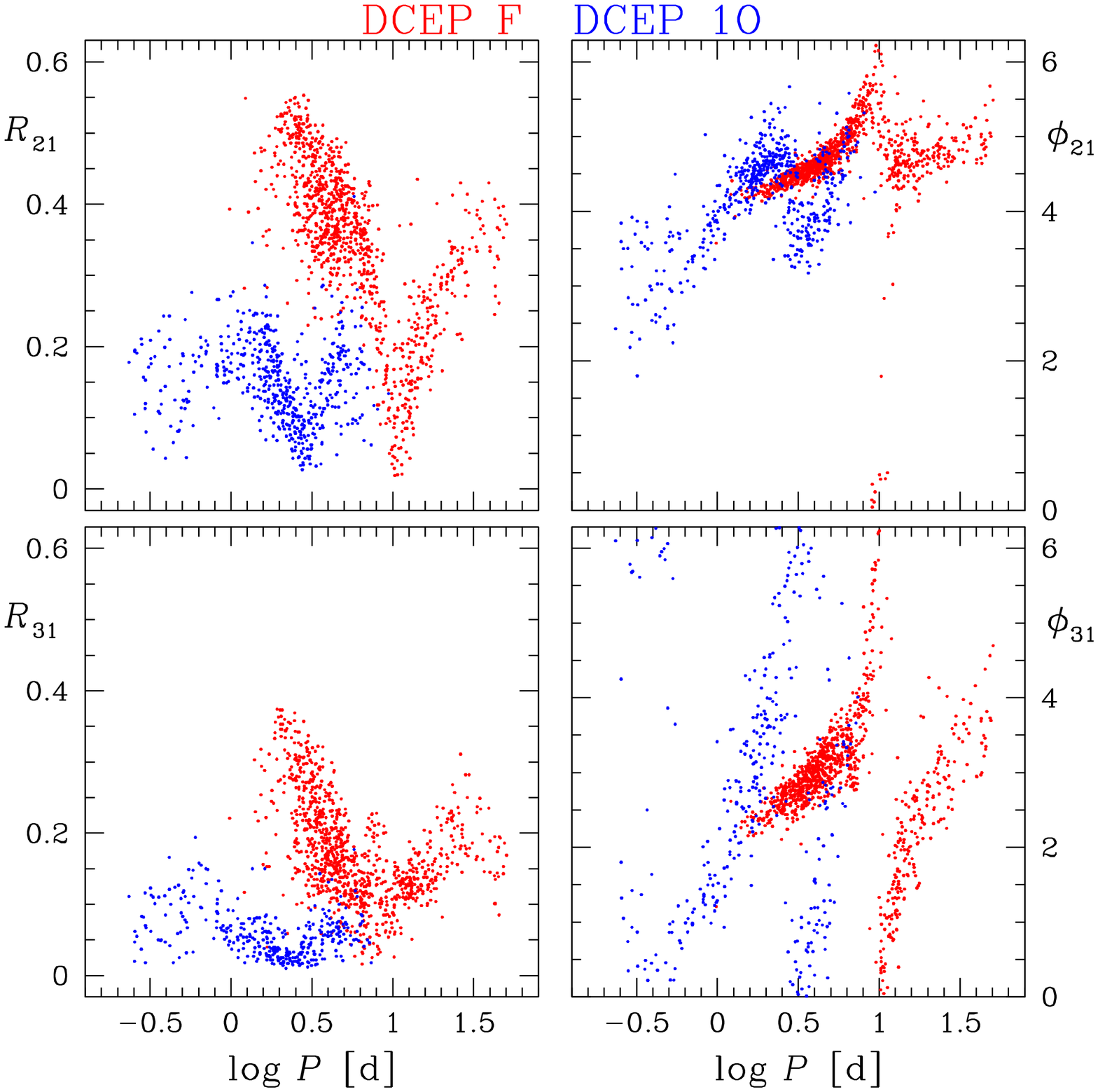}}
\FigCap{Fourier coefficient combinations $R_{21}$, $R_{31}$, $\phi_{21}$,
and $\phi_{31}$ as a function of the logarithm of the pulsation period
for fundamental-mode (red points) and first-overtone Galactic classical
Cepheids (blue points) observed by the OGLE survey. Only objects with
at least 50 $I$-band measurements and brighter than $I=18$~mag are shown.}
\end{figure}

Multi-mode Cepheids have specific period ratios. This is illustrated
in the Petersen diagram in Fig.~7, in which we plot the shorter-to-longer
period ratio in the function of logarithm of the longer period for
classical Cepheids from the Milky Way, LMC, and SMC. The points
form characteristic sequences for various period ratios. The sequence
representing 1O+2O pulsators is tight, in contrast to the sequence for
F+1O pulsators which is strongly metal-dependent. This is well seen when
one compares stars from the three galaxies. The sequences shift with
the metallicity. In general, the period ratios are higher
for stars in more metal-poor environments at a given period.

\begin{figure}[htb!]
\centerline{\includegraphics[angle=0,width=120mm]{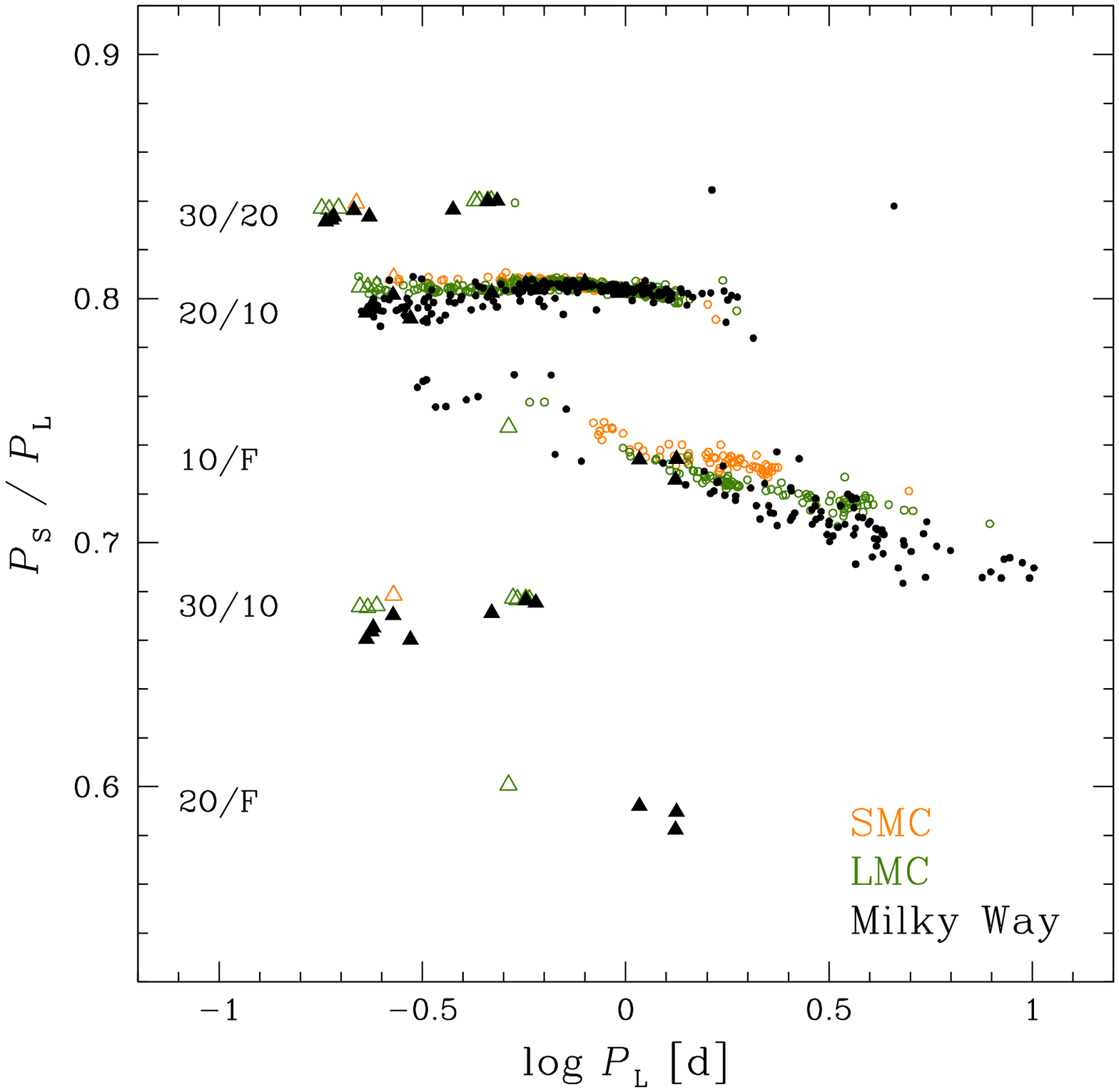}}
\FigCap{Petersen diagram with the positions of multi-mode radially pulsating
classical Cepheids from the Milky Way (black filled symbols), LMC (dark green
open symbols) and SMC (open orange symbols). Circles represent double-mode
pulsators (F+1O, 1O+2O, and 2O+3O), while triangles represent triple-mode
pulsators (F+1O+2O and 1O+2O+3O). Location of the sequences of stars
clearly depends on the metal content of the environment.}
\end{figure}

In Table~2, we provide information on numbers of Galactic classical
Cepheids pulsating in various modes and their combinations. Single-mode
pulsators is the most common group of stars (91.4 per cent), while
triple-mode pulsators are very rare (0.3 per cent). In the whole
set of variables, fundamental-mode pulsators constitute 63.8 per cent.
A detailed comparison between the collections of classical Cepheids
from the Milky Way, M31, and Magellanic Clouds is presented in Table~3.
If the real fraction of first-overtone Cepheids is slightly higher
than one-third, as observed in the Magellanic Clouds, then many such
pulsators in our Galaxy are yet to be discovered. Observations of the
Magellanic Clouds also indicate that second-overtone Cepheids are not
extremely rare and more of them should exist in the Milky Way.
Very recently, Rathour \etal (2021) performed a frequency analysis of
OGLE-IV photometry for Galactic classical Cepheids and detected
possible additional radial and non-radial modes in nearly forty
pulsators. However, the presence of new frequencies requires confirmation
once new brightness measurements are collected\footnote{Due to the
Covid-19 pandemic, OGLE stopped regular observations on 18 March 2020.}.

\begin{table}[htb!]
\centering \caption{\small Census of Galactic classical Cepheids pulsating
in various modes and their combinations: fundamental mode (F), first
overtone (1O), second overtone (2O), and non-radial 0.62-mode (X)}
\medskip
{\small
\begin{tabular}{lr}
\hline
Mode(s)    &  Number of stars \\
\hline
F          &  2140 \\
1O         &   924 \\
2O         &     1 \\
F+1O       &    93 \\
1O+2O      &   180 \\
1O+X       &     1 \\
2O+3O      &     2 \\
F+1O+2O    &     3 \\
1O+2O+3O   &     8 \\
\hline
Total      &  3352 \\
\hline
\end{tabular}}\\
\end{table}

\begin{table}[htb!]
\centering \caption{\small Incidence rates for known classical Cepheids
in four Local Group galaxies}
\medskip
{\small
\begin{tabular}{lrrrrr}
\hline
Galaxy     & Total &  F~~~~~(\%) &  1O~~~~(\%) & 2O~~~~(\%) & multi-mode (\%)\\
\hline
Milky Way  &  3352 & 2140 (63.8) &  924 (27.6) &   1 (0.03) & 287 (8.6) \\
M31        &  1969 & 1662 (84.4) &  307 (15.6) &   ?        &   ? \\
LMC        &  4706 & 2477 (52.6) & 1776 (37.7) &  26 (0.55) & 427 (9.1) \\
SMC        &  4944 & 2754 (55.7) & 1791 (36.2) &  91 (1.84) & 308 (6.2) \\
\hline
\end{tabular}}\\
\end{table}

Below, we provide some interesting records referring to the Galactic
classical Cepheids.

\begin{itemize}
\item $\beta$~Dor has the highest absolute galactic latitude ($b=-32\zdot\arcd77$).
\item The brightest classical Cepheid in the sky is Polaris.
\item S~Vul is a fundamental-mode pulsator with the longest period of 68.651~d.
\item OGLE-GD-CEP-1628 is a first-overtone pulsator with the longest period of 9.437~d.
\item V473~Lyr is the only known single-mode Cepheid pulsating in the second overtone.
\end{itemize}


\Section{Purity and Completeness of the List}

It is difficult to determine purity of a list of objects constructed
based on various catalogs. To estimate it, we can use the most reach
and uniform source of Galactic classical Cepheids, the OGLE collection.
We found that 10 detections out of 1973 stars were in fact artifacts
in the vicinity of much brighter real Cepheids saturated in OGLE-IV images.
Additional 47 objects were noted as uncertain cases in Soszy\'nski \etal (2020).
This means that 1916 variables or 97.1 per cent of the sample are
{\it bona fide} classical Cepheids. The purity of Cepheids pulsating in the
fundamental mode solely is even higher, 1179/1201 or 98.2 per cent.
The OGLE collection contains on average fainter objects than those from
other optical catalogs, thus the estimated values can be treated as lower limits.

We decided to include in our list only classical Cepheids confirmed in the
optical range. Verification of candidates for Galactic Cepheids from
near-infrared VVV observations showed that more than a half of the objects
are of other variability type. Out of 689 candidates for classical Cepheids
in D\'ek\'any \etal (2019), 197 stars are present in OGLE images,
of which 136 objects show periodic variations in the optical range.
It turns out that only 63 variables are real classical Cepheids.
The remaining variables are type II Cepheids (5 stars), rotating (23 stars)
and eclipsing variables (17 stars), and stars of uncertain variability type (28).
Accordingly, the purity of the near-infrared observations is only
46.3 per cent. The reason for this very low success rate is
that near-infrared light curves of pulsating stars have smoother,
nearly-sinusoidal shapes, on average lower amplitudes, and more often
can be confused with variables of other types. Large number of
measurements and long-term stability of the light curve are required
to properly classify the variables as classical Cepheids.

It is even more complicated to assess completeness of the presented list of
classical Cepheids. The OGLE mosaic camera covers about 93 per cent of each
OGLE-IV field. The remaining area are gaps between the CCD detectors. From
the brightness distributions presented in Fig.~3 (drawn in red), we see that
the census of Galactic fundamental-mode classical Cepheids is highly complete
down to $G=18$ mag. In mildly reddened regions (blue histograms in Fig.~3),
practically all Cepheids are known. However, Fig.~4 shows that there
are missing objects in the period distributions. There is a deficit of
about 100 fundamental-mode pulsators with periods around 10 days
and about 50 first-overtone pulsators with periods around 2.5 days.
This is about 5 per cent of the F and 1O samples. Moreover, we cannot
exclude that there are yet undiscovered low-amplitude Cepheids
($<0.1$ mag), similar to Polaris, that are entering or exiting the main
instability strip. Several relatively bright ($I<12.5$ mag) unknown
classical Cepheids are expected to exist in the nearby Sagittarius arm
since OGLE has conducted only a deep monitoring of the inner Galactic
bulge. Overall, we can estimate that the completeness of the presented list
of Cepheids is of about $0.95\times0.93$ or 88 per cent down to $G=18$ mag.
This result refers to stars observed in the optical range, while hundreds of
unknown classical Cepheids likely reside at the far side of the Milky Way's
disk hidden behind thick clouds of interstellar dust along spiral arms.


\Acknow{
We thank Prof. Nikolai N. Samus for sending us information on Cepheids
in the first GCVS edition. This research has been supported by the National
Science Centre, Poland, grant MAESTRO 2016/22/A/ST9/00009 to I.S.

We used data from the European Space Agency (ESA) mission {\it Gaia},
processed by the {\it Gaia} Data Processing and Analysis Consortium
(DPAC). Funding for the DPAC has been provided by national institutions,
in particular the institutions participating in the
{\it Gaia} Multilateral Agreement.}


\end{document}